\documentclass[doublecol]{epl2} 
\RequirePackage{amssymb} 

   \title{Effect of topology on dynamics of knots in polymers under tension}
\shorttitle{Effect of topology on dynamics of knots in polymers under tension} 

\author{R. Matthews \and A.A. Louis  \and J.M. Yeomans}
\shortauthor{R. Matthews \etal}

\institute{                    
  \inst{1} Rudolf Peierls Centre for Theoretical Physics, 1 Keble Road, Oxford 0X1 3NP, England\\
}


\abstract{We use computer simulations to compare the dynamical behaviour of torus and
even-twist knots in polymers under tension. The knots diffuse through a mechanism similar to reptation. Their friction coefficients grow linearly with average knot length for both knot types. For similar complexity, however, the torus knots diffuse faster than the even twist knots. The knot-length auto-correlation function exhibits a slow relaxation time that can be linked to a breathing mode. Its timescale depends on knot type, being typically longer for torus than for even-twist knots. These differences in dynamical behaviour are interpreted in terms of topological features of the knots.}
\begin{document}
\bibliographystyle{eplbib}
 
\maketitle

The scientific study of knots has a long history, dating back at least to Johann Friedrich Gauss in the early 19th century. More recently, deep connections between knot theory, statistical mechanics~\cite{jones} and quantum field theory~\cite{witten} have stimulated a great deal of research both in physics and mathematics.    

The discovery of knots in bacterial DNA~\cite{liu} and proteins~\cite{mansfield,kolesov,dzubiella}, as well as earlier work on synthetic polymers~\cite{frisch} helped broaden the relevance of knot theory to chemistry and biology. Indeed, knots and links can be introduced {\em in vivo} into cellular DNA through processes such as replication~\cite{olavarietta} and are regulated through enzymes such as type II topoisomerases that can both knot and unknot DNA~\cite{watt}. Biologists have exploited these topological effects to make many discoveries about the nature of DNA in cells~\cite{wasserman}. DNA is thought to be highly knotted inside some viral capsids~\cite{arsuaga}, and we have recently suggested that these knots may control the ejection speed of a bacteriophage's DNA~\cite{matthews}.  

The discovery of knots in nature raises further questions about the dynamics of knotted polymers. In contrast to unknotted polymers, where dynamic behaviour is fairly well understood~\cite{doi}, many basic questions remain open. In polymers without tension, a long topological timescale, not predicted by the standard Rouse model~\cite{doi}, was first discovered in the relaxation of the radius of gyration for knotted polymers~\cite{quake}. Later work showed that this topological relaxation time decreased with increasing knot complexity~\cite{lai}, and could also be observed by measuring the knot length autocorrelation function~\cite{orlandini}. Related work~\cite{yu,lai2} showed that, by contrast, this  timescale increased with knot complexity when knotted ring polymers are cut. 

Knots can be artificially introduced into biopolymers such as actin filaments~\cite{arai} and linear DNA ~\cite{bao} with single molecule techniques. Bao $et.$ $al.$\cite{bao} measured the diffusion coefficients of DNA knots by fluorescence microscopy. The knot friction coefficient grew roughly linearly with knot length, so that more complex knots diffused slower. Similar diffusion coefficients were obtained with simulations by Vologodskii~\cite{volo}. In these studies polymers were under tension. It has also been argued that knots could exist in metastable tight states due to entropy alone~\cite{grosberg}.
 
The experiments performed in ref.~\cite{bao} utilized tensions $0.1$-$2$ $pN$, which may be in a similar range to forces induced {\em in vivo} by enzymes such as polymerases~\cite{liu2}.  Huang and Makarov~\cite{huang} showed that the experiments were in the intermediate ``elastic" regime where the external force $f > k_B T/l_p$, with $l_p$ the persistence length. Here the force aligns the segments of the chain in the general direction of the force,  and the knot size is determined by the bending elasticity of the chain vs the force. The diffusion coefficient is only weakly dependent on tension, an effect also observed in experiments~\cite{bao}. The elastic regime is bounded from above by a  ``tight knot" regime where $f \gg k_BT /l_p$ and the knot properties are dominated by details of molecular interactions between monomers. At much lower tensions $f \ll k_B T/l_p$ in the ``blob" regime, the knot size fluctuates widely.

Most previous work on knot dynamics has focussed on the ``blob" regime. However, experiments on artificially knotted DNA~\cite{bao}, as well as the expectation that some knots {\em in vivo} may be held under tension~\cite{matthews,liu2}, mean it is important to explore the effect of topology on dynamics in the elastic regime. This task will be our main focus here. 

\begin{figure}[ht!]
\includegraphics[scale=0.3]{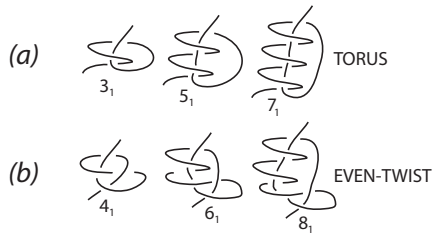}
\caption{\label{fig:knot_types}(a) The first three torus knots. (b) The first three even-twist knots.}
\end{figure}

We focus on two classes of knot topologies: torus knots and even-twist knots. The first few knots in each group are shown in Fig.~\ref{fig:knot_types} - knots of higher complexity have more crossings but a similar structure. We use standard notation, $C_k$, where $C$ denotes the minimal number of crossings in a projection onto a plane, and $k$ distinguishes between knots with the same number of crossings~\cite{orlandini2}. 

To study the behaviour of  knots in  polymers under tension we used a bead-spring model coupled to a coarse grained solvent~\cite{malev2,matthews,ripoll}, briefly described here. The beads interact via the potential:

\begin{eqnarray}\nonumber
\beta V&=&4\beta \epsilon\sum_{j>i}\sum_{i}\left[\frac{\sigma}{\mid\vec{r}_i
-\vec{r}_{j}\mid}\right]^{12}
\\&-&\frac{k R_0^2}{2}\sum_{i}\ln\left[1-\left(\frac{\mid\vec{r}_i
-\vec{r}_{i-1}\mid}{R_0}\right)^2\right]
\label{potential}
\end{eqnarray}

where the first term is an excluded volume interactions and the second term is a FENE spring potential to simulate bonds. The parameters were chosen as: $\beta \epsilon = 1$, $\sigma = 1$, $k = 30$ and $R_0 = 1.5$.  The equations of motion were integrated with a velocity Verlet algorithm. 

In this paper we used flexible polymers as we are mainly interested in generic, qualitative effects. For a quantitative comparison to experiments, it would be important to also include a bending potential that captures the semi-flexible nature of DNA, as has been done in previous simulations ~\cite{volo,huang}. For example, in the elastic regime we study, semi-flexibility can lead to knots that extend fewer persistence lengths~\cite{bao} than those in our simulations.

The polymer was coupled to a mesoscopic solvent modelled with stochastic rotation dynamics (SRD)~\cite{malev}.  On average there are $5$ solvent particles per $\sigma^3$, and they provide a thermostat which conserves momentum and so preserves hydrodynamic interactions between monomers. The model thus describes a flexible polymer undergoing Brownian motion in a hydrodynamic solvent~\cite{malev2,ripoll}.

\begin{figure}[ht!]
 \includegraphics[scale=0.35]{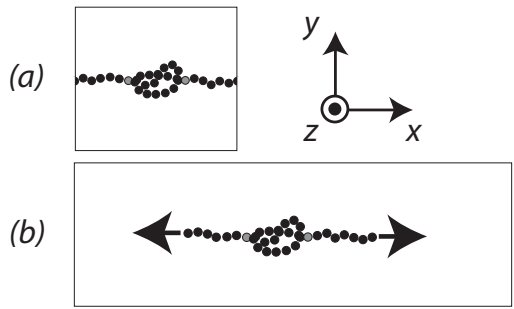}
\caption{\label{fig:set_up}The two geometries used in the simulations. In  (a) the polymer ends are connected  using periodic boundary conditions, forming a ring polymer under tension.  In (b), the polymer is linear and the two end beads are subjected to a constant force to provide the tension. Grey circles indicate the beads identified as the first and last beads of the knot by the knot tracking algorithm}
\end{figure}

To simulate a polymer under tension, two different geometries, illustrated in Fig.~\ref{fig:set_up}, were considered. In the geometry (a), a simulation box of 32$\sigma$ by 32$\sigma$ by 20$\sigma$ or 21$\sigma$ was used. The polymer was connected to itself across the periodic boundary.

By varying the number of beads in the polymer, the tension can be changed. This approach offers the advantage of no free ends. Simulations may be run as long as necessary for good statistics without worrying about unknotting or other end effects. In the geometry (b), a much larger simulation box of 32$\sigma$ by 32$\sigma$ by 300$\sigma$ was used and the polymer ends were not joined. Instead the first and last beads were subjected to a constant force. Geometry (b) was mainly used to test the reliability of the more efficient geometry (a).

Knots were introduced by hand. The first and last beads of a knot, shown in Fig.~\ref{fig:set_up}, can be identified by finding bond crossings of bead to bead vectors from the projection in the x-y plane (see Fig.~\ref{fig:set_up}). The midpoint between the first and last beads was taken as the knot's position and the difference as its length. Distances were measured along the polymer contour, as in the experiments~\cite{bao}. 

\begin{figure}[ht!]
\includegraphics[scale=0.22]{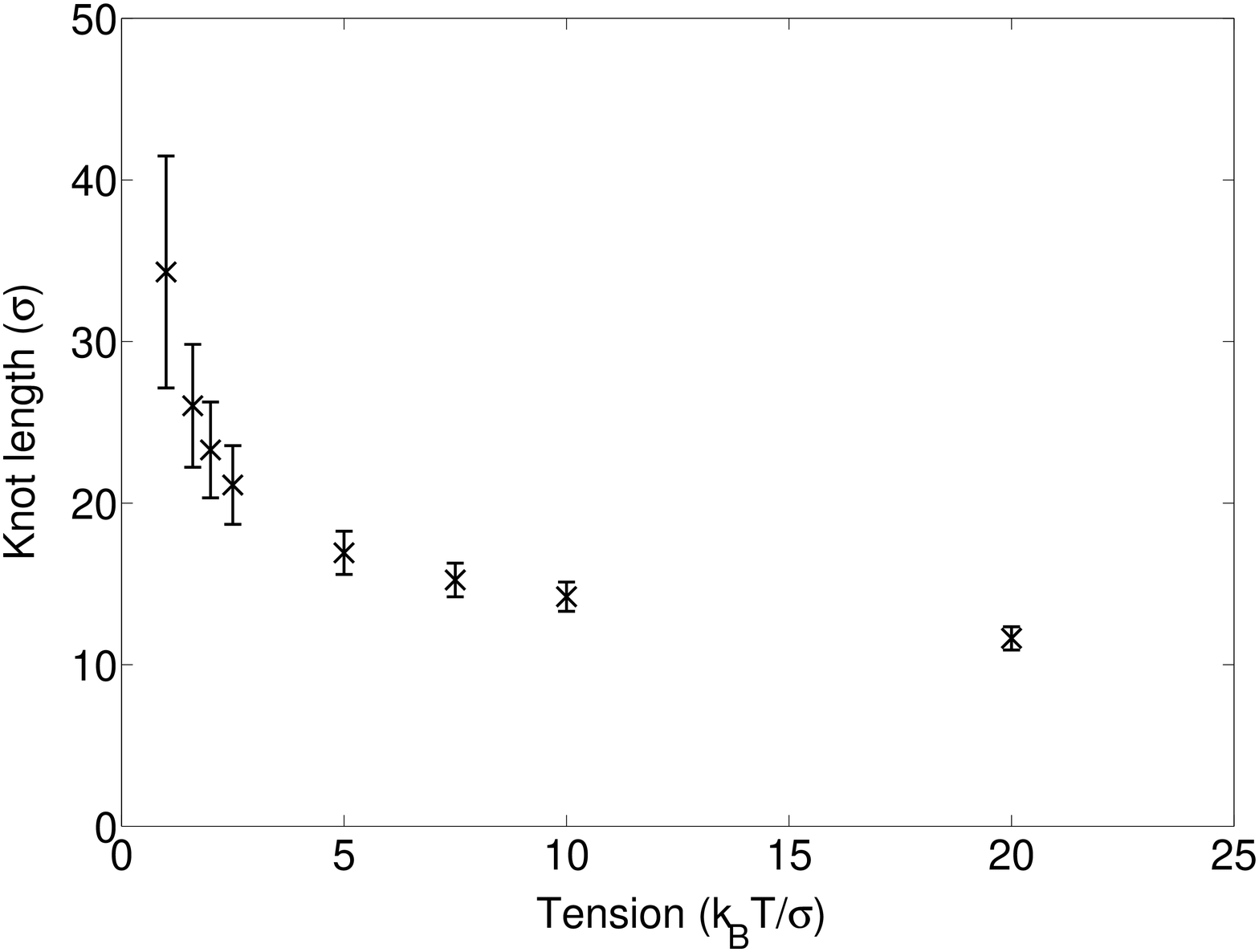}
\caption{\label{fig:three_length_ten}Average knot length for the $3_1$ knot on a polymer under tension. Simulations are done with geometry (b).}
\end{figure}

Fig.~\ref{fig:three_length_ten} shows that the  length of the $3_1$ knot  decreases with tension and plateaus at higher tension, consistent with ref~\cite{huang}.  This behaviour holds for other knots too.  We will primarily use a tension of $f= 5 \frac{k_B T}{\sigma}$, in the elastic regime since for a flexible polymer $\sigma \sim l_p$.  If we were modeling DNA,  where $l_p \approx 50 nm$, our force would map onto $f \approx 0.4$ pN, which is comparable to the experiments of Bao {\em et al.}~\cite{bao}, which were also in the elastic regime.

\begin{table}[ht!]
\caption{\label{tab:knot_length} Average knot length for linear polymers under a tension of $5 \frac{k_BT}{\sigma}$ for geometries (a) and (b). }
\begin{center}
\begin{tabular}{lcr}

Knot type &Length in (b) &Length in (a)\\
&geometry(${\sigma}$)&geometry(${\sigma}$)\\\hline
$3_1$&$17\pm1$&$17\pm1$\\
$4_1$&$24\pm2$&$24\pm1$\\
$5_1$&$28\pm2$&$29\pm2$\\
$6_1$&$35\pm2$&$35\pm2$\\
$7_1$&$39\pm2$&$39\pm2$\\
$8_1$&$46\pm3$&$45\pm2$\\
$9_1$&$49\pm3$&$50\pm2$\\
$10_1$&$56\pm3$&$57\pm2$\\
$11_1$&$59\pm3$&$59\pm2$\\
$12_1$&$66\pm3$&$67\pm2$

\end{tabular}
\end{center}
\end{table}

In Table~\ref{tab:knot_length} we compare the average length of different knots for polymers under a tension of  $f=5 \frac{k_BT}{\sigma}$.  The knot length increases approximately linearly with the number of essential crossings, an effect we observe at other tensions as well. For geometry (a) it is harder to set tension explicitly. Instead either the number of polymer beads was changed, or the box length was varied between $L=20 \sigma$ and $L=21 \sigma$ (the SRD algorithm uses boxes of size $\sigma$ so constrains us to integer values). By comparing the knot length to the linear geometry, a similar effective tension could be simulated.  We tested that the knot diffusion coefficient only depends very weakly on tension in the range $f=2-8 \frac{k_BT}{\sigma}$ , as expected for the  elastic regime~\cite{bao,huang}.

\begin{figure}[ht!]
\includegraphics[scale=0.22]{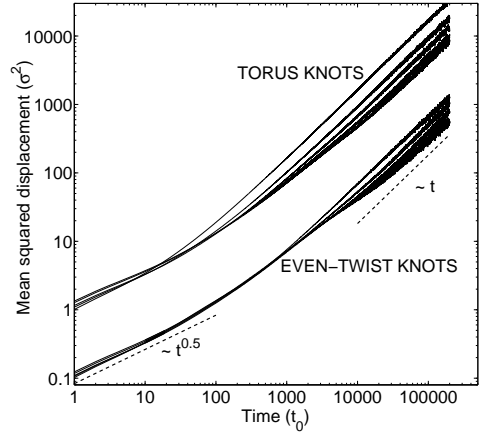}
\caption{\label{fig:all_knots_msd_log_log} Mean squared displacement curves plotted on log-log scales. The upper lines are for the torus knots and have been displaced on the vertical axis by a factor of 10. The lower lines are the even-twist knots. Within each group knot complexity increases from top to bottom. The dashed lines have slopes of 0.5 and 1.}
\end{figure}

\begin{table}[ht!]
\caption{\label{tab:time_diff_coeff}Relaxation timescales and diffusion coefficients for knots in a polymer under tension. Timescales were found by an exponential fit to the autocorrelation decay between 0.5 and 0.05, see Fig.~\ref{fig:knot_length_correl_inset}.}
\begin{center}
\begin{tabular}{lcr}

Knot &Relaxation timescale &Diffusion coefficient\\
type  &($\times10^3t_0$) &  ($\times10^{-3}\frac{\sigma^2}{t_0}$)\\\hline
$3_1$&$0.020\pm0.001$&$8.0\pm0.4$\\ 
$4_1$&$0.029\pm0.001$&$3.1\pm0.2$\\ 
$5_1$&$2.1\pm0.1$&$4.5\pm0.3$\\
$6_1$&$1.2\pm0.1$&$2.5\pm0.2$\\ 
$7_1$&$2.5\pm0.1$&$3.4\pm0.2$\\ 
$8_1$&$1.4\pm0.1$&$2.0\pm0.1$\\ 
$9_1$&$3.5\pm0.1$&$2.3\pm0.2$\\
$10_1$&$1.9\pm0.1$&$1.4\pm0.1$\\
$11_1$&$5.2\pm0.1$&$2.1\pm0.1$\\ 
$12_1$&$2.1\pm0.1$&$1.3\pm0.1$

\end{tabular}
\end{center}
\end{table}

We begin by studying the diffusive motion of the knots along the chain.  In Fig.~\ref{fig:all_knots_msd_log_log} we show the mean square displacement of the torus and even-twist knots as a function of time, measured in simulation units $t_0 = (\sigma/2) \sqrt{m_f/k_B T}$, where $m_f$ is the mass of a fluid particle. At shorter times, we observe sub-diffusive behaviour with  $<x^2> \sim \sqrt{t}$, but this  crosses over to Fickian diffusive motion, typically around $t \gtrsim 10^5 t_0$. The resulting diffusion coefficients, extracted from the mean-square displacements at later times (typically $t=10^5 - 2 \times 10^5 t_0$), are listed in Table~\ref{tab:time_diff_coeff}.  The diffusion coefficients  clearly  decrease with increasing crossing number, but for a given complexity, torus knots have larger diffusion coefficients than even-twist knots.  We observed a similar topology induced non-monotonic behaviour  with crossing number in simulations of knotted polymer ejection from a capsid~\cite{matthews}.

\begin{figure}[ht!]
 \includegraphics[scale=0.22]{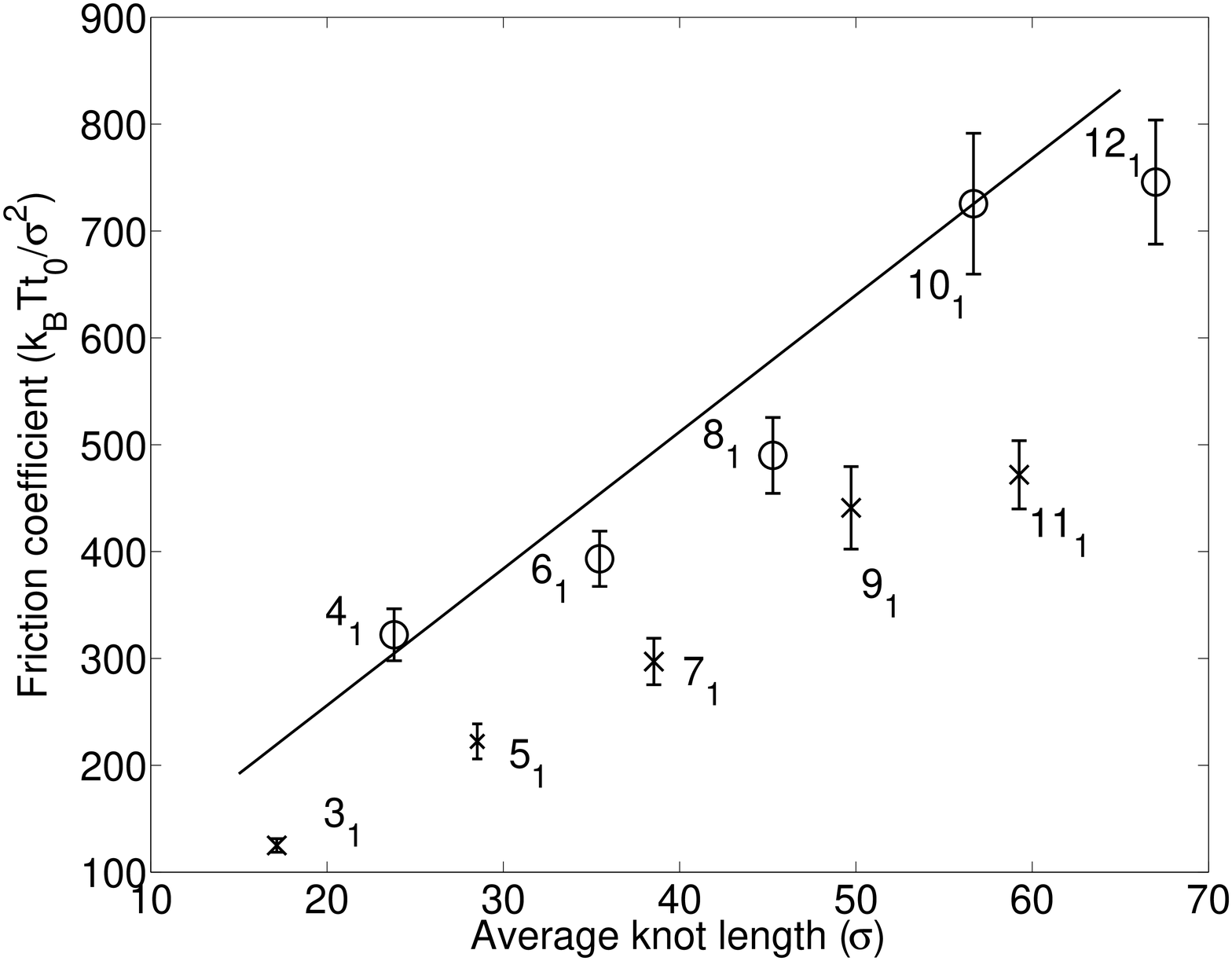}
\caption{\label{fig:fric_coeff}Friction coefficient against average knot length for torus knots (crosses) and even-twist knots (circles). The solid line is $\zeta\times$average knot length, where $\zeta$ is the friction coefficient of a monomer in the SRD solvent (each bead contributes a length of $\sigma$.)}
\end{figure}

The diffusion coefficient can be related to the friction coefficient through the Einstein relation $\xi = k_B T/D$. We show results for friction vs.\ knot length in Fig.~\ref{fig:fric_coeff}.  Both knot groups show the same linear dependence on length, with the same slope, but there is  a constant shift between them.  A very similar linear dependence of the friction on length was measured in experiments~\cite{bao} as well as in simulations~\cite{huang}.  In both cases the only even-twist knot measured was $4_1$, and the friction was consistently higher than that expected from torus knots. We also plot a simple approximation to the friction, $\zeta\times$ the average knot length, where $\zeta$ is the friction coefficient of a monomer in the SRD solvent.  This simple approximation provides a good estimate of the slope of the friction vs.\ knot length curve in this tension regime, as also seen in ref.~\cite{huang}.

For unknotted polymers, neglecting the hydrodynamic interactions (Rouse approximation) leads to a  linear dependence on the number of beads, but taking the hydrodynamics into account (Zimm approximation) changes the scaling  substantially~\cite{doi}. The methods used in refs.~\cite{volo,huang} neglect hydrodynamic interactions, but the simulation technique we use here can reproduce the correct Zimm scaling for linear polymers~\cite{malev2,ripoll}.  Including hydrodynamic interactions {\em quantitatively} changes the magnitude of the knot diffusion coefficients by about a factor of 1.5 but does not appear to alter their {\em qualitative} scaling with knot length.

The difference in friction between torus and even-twist knots can be linked to an important topological distinction. As knots diffuse along the chain, they must reptate through themselves~\cite{bao}. From Fig.~\ref{fig:knot_types}, one can appreciate that the polymer may pass relatively smoothly through a torus knot, always curving in the same sense. For an even-twist knot, however, the twist at one end introduces a sharp inversion of the direction in which the polymer is curving, making the polymer passage less smooth and increasing the effective friction. Since this inversion occurs only once in an even-twist knot, it may explain why the frictions are higher than those of the torus knots by an additive constant.

Evidence that  the main mode of knot diffusion in this regime is through reptation can be found in the mean-square displacement data in Fig.~\ref{fig:all_knots_msd_log_log}. At short times the motion is sub-diffusive because as the polymer in the knot tries to move it tends to be bounced back by other sections of the knot. At longer times, when the polymer contour has relaxed within the constraints formed by the knot, we observe ordinary Fickian diffusion.  This crossover corresponds to two regimes in the reptation model~\cite{doi} where the mean squared displacement of the polymer $along$ $its$ $own$ $contour$ increases with the exponents we observe. Our results are also not inconsistent with a ``sliding knot'' picture such as in ref.~\cite{huang}. 
 
The mapping of dynamics from coarse-grained simulation units to physical units is always subtle~\cite{padding}. In the experiments of ref.~\cite{bao},  $D \approx 1.25 \mu m^2/s$ for a $3_1$ knot. Comparing to our  simulated value for the same knot, and assuming $l_p \sim \sigma$ gives  a mapping of  $t_0 \approx 16 \mu sec$.   Very similar values for $t_0$ are found when using the other knots to do the mapping.  This suggests that sub-diffusive  $<x^2> \sim \sqrt{t}$ behaviour should persist up to a timescales of order $ms$, with the full crossover to Fickian diffusion occuring at time scales on the order of $s$. These time scales correspond qualitatively to the experiments of ref.~\cite{bao}.

\begin{figure}[ht!]
\includegraphics[scale=0.22]{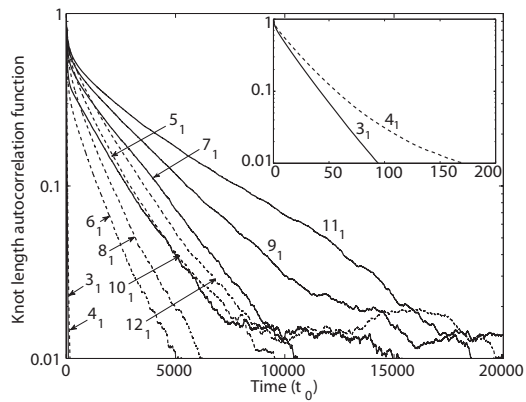}
\caption{\label{fig:knot_length_correl_inset}Decay of the knot length autocorrelation with time. Results for torus knots are plotted using solid lines and for even-twist knots using dashed lines. The inset shows $3_1$ and $4_1$ results on a different scale.}
\end{figure}

We next investigate the fluctuations and dynamic modes of the diffusing polymer knot under tension. 
Orlandini $et$ $al$~\cite{orlandini} showed that the long timescale originally observed in the radius of gyration autocorrelation function in the ``blob" regime~\cite{quake}  can also be observed in the in the decay of the knot length autocorrelation function $\frac{<l(t)l(0)> - <l(0)><l(0)>}{<l(0)l(0)> - <l(0)><l(0)>}$, where $l(t)$ is the length of the knot at a given time. We investigated the same autocorrelation function for knots under tension, and plot the results in Fig.~\ref{fig:knot_length_correl_inset}. The decay of the knot length autocorrelation function is approximately exponential, and the associated timescale (the inverse of the slope of the lines) of this exponential decay depends on the knot type, see Table~\ref{tab:time_diff_coeff}. The timescale increases with knot complexity and is also significantly longer for torus knots than for even-twist knots of similar complexity. There is a remarkably large jump (a factor of about $100$) in the decay timescale between the most simple knots ($3_1$ and $4_1$) and the more complex knots ($5_1$ and larger). Preliminary investigations of polymer length, N, dependence showed that the timescales for the more complex knots vary weakly with N, whereas the $3_1$ and $4_1$ timescales are more sensitive to N.

\begin{figure}[ht!]
\includegraphics[scale=0.3]{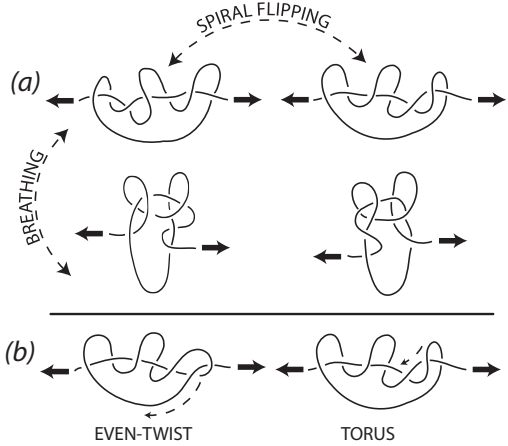}
\caption{\label{fig:knot_breath_modes_combine} (a) Examples of knot modes (for $7_1$). Arrows indicate the direction of the tension force. The change between the left and right involves inversion of the spiral - the spiral-flipping mode. The change between top and bottom involves switching the alignment of the spiral with respect to the tension force - the breathing mode. (b) Origin of the faster breathing mode in even-twist knots. Left: in even-twist knots, the end of the spiral, the twist, may slide relatively freely along the strand that passes through it. Right: if the end of the spiral moves in a similar manner in a torus knot it collides with other loops.}
\end{figure}

To identify the source of the slow decay of the knot length autocorrelation function, the knot configurations were more closely investigated. Two different modes were identified: a spiral-flipping mode and a breathing mode, both depicted in Fig.~\ref{fig:knot_breath_modes_combine}(a). 

The spiral-flipping mode involves switching between a state where the strand entering from the left predominantly curves around the strand entering from the right and vice-versa. The breathing mode involves switching between a state where the axis of the spiral is aligned with the tension force and one where the spiral forms a loop perpendicular to it. To characterise the knot state, two order parameters were used: one for each mode. 

For the spiral-flipping mode the sum of angles between successive bonds - in the plane defined by those bonds - multiplied by the displacement from the centre of the knot was taken over all of the beads in the knot. Depending on whether the polymer is more curved towards one end of the knot or the other, this quantity is positive or negative. For the breathing mode the ratio between the maximum distance between any two beads in the knot in the direction of the tension force to the maximum distance perpendicular to the tension force was taken. When the spiral of the knot is aligned with the tension this quantity is larger.

\begin{figure}[ht!]
\includegraphics[scale=0.21]{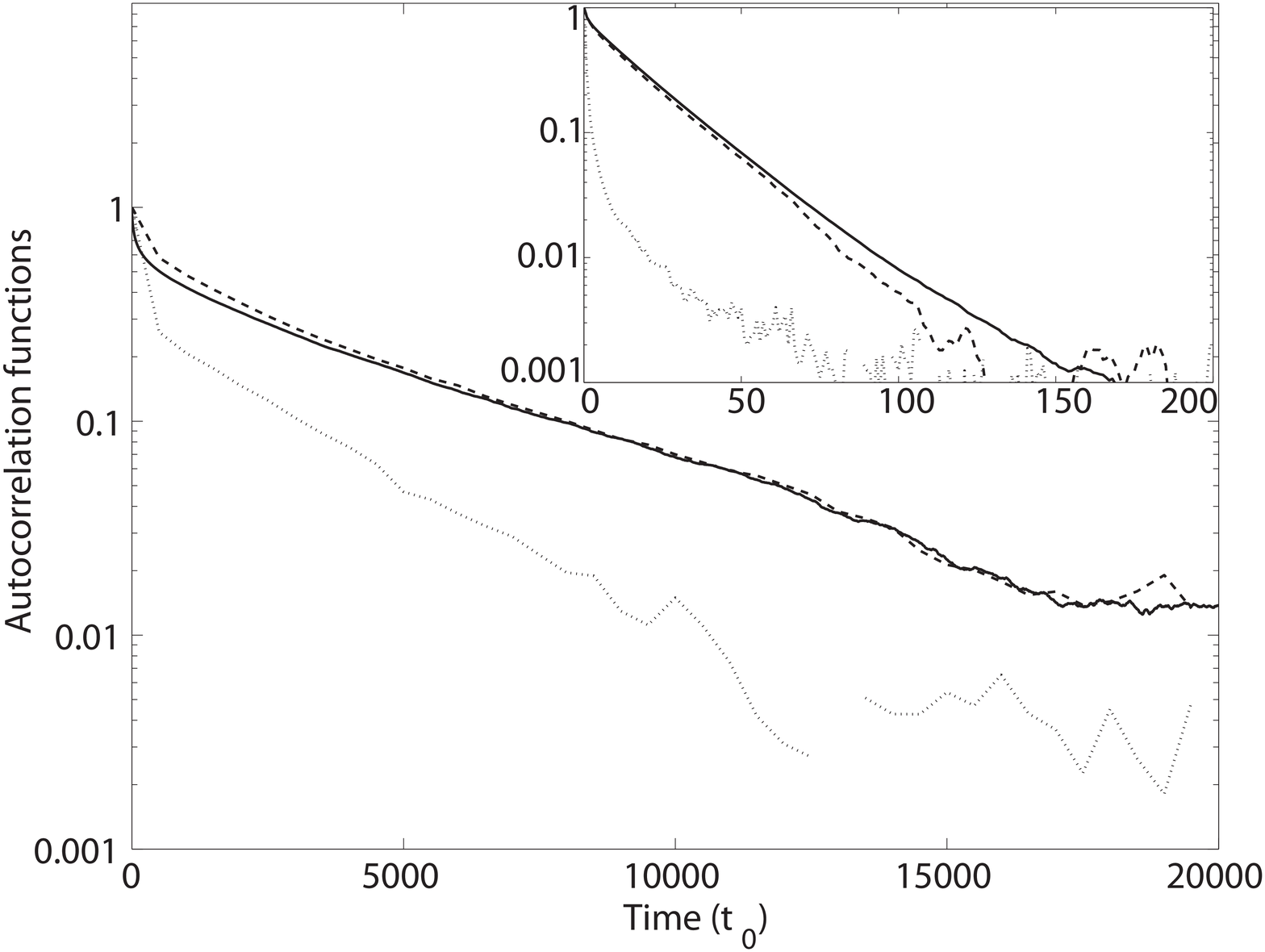}
\caption{\label{fig:three_one_eleven_one_autocorrels} Autocorrelation functions for $11_1$ for the knot length (solid line), the breathing mode order parameter (dashed line) and the spiral-flipping mode order parameter (dotted line). The inset shows results for $3_1$ on a different scale.}
\end{figure}

We calculated the autocorrelation functions, $\frac{<\phi(t)\phi(0)> - <\phi(0)><\phi(0)>}{<\phi(0)\phi(0)> - <\phi(0)><\phi(0)>}$, where $\phi$ is the order parameter value, for both order parameters. As an example, the autocorrelation functions for the $3_1$ and $11_1$ knots are depicted in Fig.~\ref{fig:three_one_eleven_one_autocorrels}. The breathing autocorrelation function has a very similar decay to the knot length autocorrelation function, whereas the spiral flipping autocorrelation function decays more rapidly. Nevertheless, both modes show a large jump in time scale between the $3_1$ and $4_1$ and the more complex knots.  
   
These results suggest that the breathing mode is responsible for the slow decay of the knot length autocorrelation. The average knot lengths for high and low values of the breathing mode order parameter support this. The extended configuration has a larger knot length; for the most complex knots a difference of about 3$\sigma$ is seen.

The coupling to the breathing mode offers an explanation for the large jump in the timescale between $3_1$ and $4_1$ and the more complex knots. The spiral of the most simple knots consists of only one loop and so shifting from a configuration which is extended in the tension direction only involves changing its shape. In contrast, for more complex knots, it involves co-ordinated movement of multiple loops. It also offers an explanation for the difference between torus and even-twist knots. As depicted in Fig.~\ref{fig:knot_breath_modes_combine}(b), an even-twist knot may deform from being extended by sliding the twist along the strand that passes through it. In a torus knot the equivalent movement is blocked by the loops of the spiral so breathing is slower.

Note that there is no big jump in the diffusion coefficients between $3_1$ and $4_1$ and the more complex knots.  While the breathing and spiral-flipping modes  may affect the crossover from sub-diffusive to diffusive behaviour (in fact the sub-diffusive range is considerably shorter for the $3_1$ and $4_1$ knots) they do not appear to determine the long-time diffusive dynamics. Further evidence follows from the fact that the torus knots have slower relaxation but faster diffusion than the even-twist knots.

It is instructive to compare our results for the longest relaxation timescale to ring polymers without tension in ref.~\cite{lai}.  By contrast with our results, the long timescale was shown to {\em decrease} with increasing knot complexity and there was no significant difference between knot groups.  This suggests that the dominant modes determining relaxation in the ``blob" regime are not related to the breathing modes we observe for polymers under tension. 

By contrast, simulations in ref.~\cite{lai2} that studied the relaxation of initially knotted polymers that are cut (without tension) exhibited a relaxation timescale that increased with knot complexity, and was longer for torus knots than for even-twist knots. The similarity is perhaps not surprising. For a cut knot to become unentangled the polymer must move along the contour formed by the erstwhile knot, a process analogous to the self-reptation~\cite{bao} of knots under tension.

In summary: we exploited a new boundary condition that minimizes end-effects to  study the effect of topology on the dynamics of  knots in polymers under tension in the ``elastic regime" where the force is strong enough to affect the behaviour of the knot, but not so strong that  details of the inter-molecular potentials dominate the behaviour.  

The observed motion of the knot, including a sub-diffusive regime of the mean-square displacement at short-times, is consistent with a self-reptation mechanism~\cite{bao}. To first order, the friction coefficient scales linearly with knot length. Topological differences also play a role. For a similar knot length or complexity, torus knots have a smaller friction (and so a larger diffusion coefficient) than even-twist knots.   

Fluctuations and dynamic modes are also affected by knot topology. The longest relaxation mode of the length-length autocorrelation function is dominated by a breathing mode. This topological timescale increases with knot complexity and is significantly longer for the torus knots than for even-twist knots of similar complexity. These differences can be rationalized by examining knot topology.

In this paper we have used a simple flexible polymer model. The main results summarized should carry over to other polymer models.  Nevertheless, it would be interesting, for example, to simulate  semi-flexible polymers that provide a better DNA model. In particular, this may give a more accurate estimate of physical timescales, and aid experiments to measure them. In addition, it would be important to study the role of tension on DNA knots {\em in vivo}\cite{liu2}.  What are the magnitudes of the biological forces, and what is  the role of macromolecular crowding on knot diffusion and relaxation rates?  Finally, the advent of nanotechnology has brought questions about the diffusion and dynamics modes of knotted polymers in strong confinement to the fore. It would be interesting to extend work that looks at the interplay of confinement and tension~\cite{metzler,moebius}.

\acknowledgements
We thank Olivier Pierre-Louis and Enzo Orlandini.


\end{document}